\documentclass[elsart12,epsf,eqsecnum]{revtex4}
\usepackage{amsmath}
\usepackage{epsfig,epsf}
\usepackage{graphicx}
\usepackage{verbatim}
\usepackage{epsfig,epsf}

\usepackage{amsmath}

\DeclareMathOperator{\arccosh}{arcCosh}

\begin{document}

\newcommand{\W} {\text{W}}
\def\beq{\begin{equation}}
\def\eeq{\end{equation}}
\newcommand{\nn}{\nonumber}

\def\beq{\begin{equation}}
\def\eeq{\end{equation}}
\newcommand{\bea}{\begin{eqnarray}}

\newcommand{\eea}{\end{eqnarray}}

\def\eq#1{{Eq.~(\ref{#1})}}
\def\fig#1{{Fig.~\ref{#1}}}


\title{Entanglement entropy  and  entropy production in the  Color Glass Condensate framework.}
\author{ Alex Kovner$^{1}$ and Michael Lublinsky$^2$}

\affiliation{
$^1$ Physics Department, University of Connecticut, 2152 Hillside
Road, Storrs, CT 06269-3046, USA \\
$^2$ Department of Physics, Ben-Gurion University of the Negev,\\
Beer Sheva 84105, Israel\\
}

\begin{abstract}
We compute the entanglement entropy of soft gluons in the wave function of a fast moving hadron and discuss its basic properties. We also derive the expression for entropy production in a high energy hadronic collision within the Color Glass formalism. We show that long range rapidity correlations give negative contribution to the production entropy. We calculate the (naturally defined) temperature of the produced system of particles, and show that it is proportional to the average transverse momentum of the produced particles.
\end{abstract}

\date{\today}

\maketitle


\mbox{}

\pagestyle{plain}

\setcounter{page}{1}




\date{\today} 

\section{Introduction}

High energy hadronic scattering is of great current interest. The structure of the hadronic wave function has been the subject of study in the recent years with the aim of better understanding possible manifestations of gluon saturation\cite{GLR}. 
The evolution of this wave function with energy, which drives the evolution of high energy scattering amplitude is encoded in the so called JIMWLK equation\cite{jimwlk,cgc,Bal}.

The stand out property of this Color Glass Condensate (CGC) wave function is that the soft gluons state``grows'' out of that of the valence modes. More specifically the wave function of the soft modes is determined by the color charge density of the valence part of the wave function. This suggests a certain ``order'' in the system, since the soft gluons are entangled with the valence ones. The measure of (dis)order in a quantum system is entropy. In particular entanglement entropy measures the degree of entanglement between different subsets of degrees of freedom in a quantum state. Entanglement entropy measures both how far the system is from a pure state, but also how close it is to a thermal one. It is therefore interesting to ask what is the entanglement entropy of a CGC wave function. 

Although entanglement entropy is a characteristic of a hadronic wave function, it also indirectly carries some information about the structure of the final state in collision of this hadron with a hadronic target. For example recently we have shown that Bose-Einstein correlations present in an incoming CGC state\cite{Bose}  manifest themselves as ridge type correlations between particles produced in the final state of an energetic collision. The CGC based calculation of such correlations\cite{DV} provides a possible explanation of the ridge correlations observed by the LHC experiments in high multiplicity p-p and p-Pb events\cite{CMS,CMS:2012qk, Abelev:2012ola, Aad:2012gla}.

In this paper we calculate the entanglement entropy of the soft modes in the CGC wave function. By soft we mean the gluon field modes with the longitudinal momentum in the ``last'' rapidity bin $\eta<\Delta Y\sim 1/\alpha_s$ which are still energetic enough to participate in scattering at relevant energy. 
A similar question has been addressed in \cite{Peschanski}, where a definition of dynamical entropy has been proposed. In the present paper we do not use an ad hoc definition, but rather directly calculate the standard entanglement entropy of a quantum state.  

We also consider the entropy of the produced system of soft particles. An earlier attempt in this direction is presented in\cite{Kutak}, while a different approach to the entropy production problem can be found in \cite{KharzeevT,CKS}  (see also the review \cite{Muller} and references therein). 
 The knowledge of the outgoing wave function in the CGC formalism allows us to directly calculate the entropy of the final state. We show that the correlations between produced particles give a well defined contribution to this entropy. This contribution is negative in accordance with one's naive expectation that stronger correlations mean a more ordered state. We define in a natural way the temperature of the produced system of particles. We show that in the weak field limit the temperature is given by $T=\pi\,\langle k_\perp\rangle/2$, where $\langle k_\perp\rangle $ is the average transverse momentum of the produced particles.

\section{The density matrix and Renyi entropy}
At high energy a hadronic wave function has a large soft gluon component. These are the softest gluons in the wave function which are energetic enough to scatter on a hadronic target. They occupy the rapidity interval $0<\eta<\Delta Y$ with $\Delta Y\sim1/\alpha_s$.  In what follows we will calculate the entanglement entropy of this component of the wave function.

In the CGC approach the hadronic wave function has the form \cite{kl1}
\begin{equation}
\Psi[a,A]=\psi[A]\chi[a,\rho]
\end{equation}
where $a$ is the soft gluon field modes, $A$ are the valence modes (with rapidities $\eta>\Delta Y$) and $\rho^a(x)$ is the color charge density due to the valence gluons. As long as the color charge density is not too large a good approximation to the soft part of the wave function is a simple coherent state

\beq\label{soft}
\chi[a,\rho]=\exp\left\{i\int _k
 b^i_a(k)\left[a^{\dagger i}_a(k)+a^i_a(-k)\right]\right\}\vert 0\rangle,
\eeq
with the Weizs\"acker-Williams field $b^i_a(k)=g\rho_a(k)\frac{ik^i}{k^2}$.
The creation and annihilation operators entering the above equation are the gluon operators integrated over rapidity,
\beq
a^a_i(k)\equiv\frac{1}{\sqrt {\Delta Y}}\int_{\eta<\Delta Y} \frac{d\eta}{2\pi} \  a^a_i(\eta,k).
\eeq
This structure is typical of the Born-Oppenheimer approximation, where the wave function of the fast degrees of freedom (soft gluons) is determined by the background of slow degrees of freedom (valence gluons).
The valence part of the wave function depends on the energy of the process, or in the present context on the total rapidity by which the hadron has been boosted from the rest frame, and is subject to JIMWLK evolution\cite{jimwlk}. At any fixed energy a valence  observable that depends only on the color charge density $O[\rho]$ is calculated as
\begin{equation}
\langle O\rangle=\int D[\rho]W_Y[\rho]O[\rho]
\end{equation} 
The rapidity dependence of $W_Y$ is determined by JIMWLK evolution. 
In this paper we will not insist that $W[\rho]$ solves JIMWLK equation, but instead will consider the (somehwat generalized)  McLerran-Venugopalan model\cite{mv}
\beq
W_P[\rho]={\cal N} e^{-\int_k
\frac{1}{2\mu^2(k)}\, \rho_a(k)\rho_a(-k)}
\eeq
where ${\cal N}$ is a normalization factor.

Note that the soft wave function eq.(\ref{soft}) depends on a single longitudinal  degree of freedom, i.e. the gluon field mode integrated over rapidity in the interval $\Delta Y$. Only this mode is relevant to our discussion in this paper. Rapidity dependence becomes significant only when the rapidity interval considered as soft becomes large enough $\Delta Y> 1/\alpha_s$. For such large rapidity intervals additional rapidity dependence of the wave function appears and our discussion would have to be amended. We will not consider this in the present paper.

The reduced density matrix of the soft gluons in the McLerran - Venugopalan model is \cite{Bose}
\begin{eqnarray}
 \hat \rho={\cal N}\int D[\rho] \;  e^{-\int_k
 \frac{1}{2\mu^2(k)}\rho_a(k)\rho_a(-k)}  e^{i\int_q
 b^i_b(q)\phi^i_b(-q)}\vert 0\rangle\langle 0\vert \; e^{-i\int_p
 b^j_c(p)\phi^j_c(-p)}\nonumber
\end{eqnarray}
where we have defined $
\phi^i_a(k)=a^i_a(k)+a^{\dagger i}_a(-k)$.
The integral over the charge density $\rho$ can be performed with the result 
\beq
\hat\rho = \sum_n\frac{1}{n!} e^{-\frac{1}{2} 
 \phi_i M_{ij}\phi_j} \, \Big[\prod_{m=1}^n M_{i_m j_m} \phi_{i_m}\vert 0\rangle\,\langle 0\vert \phi_{j_m}\Big] \, e^{-\frac{1}{ 2} 
 \phi_i M_{ij}\phi_j}
\label{rho1}
\eeq
Here we have introduced compact notations:
\beq
\phi_i\equiv \left[a^{\dagger a}_i(x)+a^a_i(x)\right]\,;\ \ \ \ \ \ \ \ \ \ \ 
\,M_{ij}\equiv \,\frac{g^2}{4\pi^2} \int_{u,v} \mu^2(u,v) \frac{(x-u)_i}{(x-u)^2}\frac{(y-v)_j}{(y-v)^2}\delta^{ab}
\eeq
Here $M$ bears two polarization, two color, and two coordinate indices, collectively denoted as $\{ij\}$.
In eq.(\ref{rho1}) summation over discrete and integration over continuous indices is implied. 

Our goal is to calculate the Von Neumann entropy of the reduced density matrix eq.(\ref{rho1}). As a warm-up exercise
we first compute the Renyi entropy. To this end we have to evaluate $tr[\hat\rho^2 ]$:
\beq
tr[\hat\rho^2 ]= \sum_{n,n^\prime}\frac{1}{n!n^\prime !}  \langle 0\vert e^{ - \phi_i M_{ij}\phi_j} \, 
\Big[\prod_{m=1}^n \prod_{m^\prime=1}^{n^\prime} M_{i_m j_m} M_{i_{m^\prime}{j_{m^\prime}}} \phi_{j_{m^\prime}}
 \phi_{i_m}\vert 0\rangle\,\langle 0\vert \phi_{j_m}\phi_{i_{m^\prime}}\Big] \, e^{- \phi_i M_{ij}\phi_j}\vert 0\rangle
\label{rho2}
\eeq
Computation of the matrix elements is straightforward as it involves calculation of two vacuum matrix elements, which can be performed by summing over all possible Wick contractions.
The explicit form of the light front vacuum wave function is
\beq \label{vac}
\langle\phi|0\rangle=Ne^{-\frac{\pi}{2} \phi_i\phi_i}
\eeq
As this is a Gaussian, it is convenient to introduce 
\beq\label{G}
G_{ij}\equiv\langle 0\vert e^{  -\phi_i M_{ij}\phi_j} \phi_i\phi_j\vert0\rangle/\sqrt{det[\pi/(\pi+2M)]}\,=\, [\pi+2M]^{-1}_{ij}
\eeq 
where we use the shorthand notation $
\pi\equiv \pi\delta_{ij}\delta^{ab}\delta^2(x-y)$.
The matrix elements in eq.(\ref{rho2}) do not vanish only when the total number of $\phi$ insertions in each one is even, that is 
\beq
n+n^\prime=2m
\eeq
The result for $tr[\hat\rho^2 ]$ is a sum of all possible contractions of $MG$
weighted with combinatoric factors. Given that both $M$ and $G$ are diagonal in color, the expansion can be organized in terms of color loops:
\beq
L_1\equiv tr[MG]\,; \ \ \ \ \ \ \ \ \ 
L_2\equiv tr[MGMG]\,; \ \ \ \ \ \ \ \ \ 
L_4\equiv tr[MGMGMGMG]\,....
\eeq 
Each trace over color gives a factor $N_c^2$. 
Performing  the averaging over the vacuum state we obtain
\beq
tr[\hat\rho^2 ]= det\left[\frac{\pi}{ \pi+2M}\right]\,\sum_m \sum_{n=0}^{2m}\frac{1}{n!(2m-n) !}\Big[ L_2^{m} (2m-1)!! \,+\, L_2^{m-2}L_4 6(2m-5)!!\,+\,\cdots] 
\eeq
Using the identity 
\beq
\sum_{n=0}^{2m}\frac{1}{n! (2m-n)!} (2m-1)!!= \sum_{n=0}^{2m}\frac{1}{n! (2m-n)!} \frac{(2m-1)!}{2^{m-1} (m-1)!}=\frac{2^m}{ m!}
\eeq
we find that the summation  over $m$ exponentiates:
\beq
tr[\hat\rho^2 ]= \exp\left[\,-tr[\ln(1+\frac{2M}{\pi})]\,+2\,L_2\,+\,\cdots\right] \,\simeq\, \exp\left[\,-2L_1\,+\,L_2\,+\,\cdots\right]
\eeq
After a more careful examination we  find that the other terms exponentiate as well
 \beq
tr[\hat\rho^2 ]= \exp\left[\,-tr[\ln(1+\frac{2M}{\pi})]\,+\,\sum_n \frac{2^{2n-1}}{ n} L_{2n}\,\right] \,=\, 
\exp\left[ - tr[\ln(1+\frac{2M}{\pi})]\,-\,\frac{1}{2}tr[\ln(1 -(2MG)^2)]  \right]
\eeq
 Finally using the definition eq.(\ref{G}) we arrive at the closed-form expression
\beq
tr[\hat\rho^2 ]= \exp\left[-\,\frac{1}{ 2}\, tr \left[\ln(1+\frac{4M}{\pi})\right]\right] \eeq 
from which the Renyi entropy is found as
\beq
\sigma_2\equiv - \ln  tr[\hat\rho^2 ]\,=\, \,\frac{1}{ 2}\, tr \left[\ln(1+\frac{4M}{\pi})\right]
\eeq
 
 \section{Von Neumann Entropy}
 
 We are now ready to compute the Von Neumann entropy of the reduced density matrix defined as 
  \beq
 \sigma^E=-tr[ \hat\rho \,\ln \hat \rho]
 \eeq
The following identity is very useful for this purpose
 \beq
\ln \hat\rho= \lim_{\epsilon\rightarrow 0}\frac{1}{ \epsilon}\,(\hat \rho^\epsilon-1)
\eeq
We first compute  $tr[\hat\rho^N]$, for arbitrary $N$, and then take the limit $N\rightarrow 1+\epsilon$. 

Consider the generalization of eq. (\ref{rho2}) to the calculation of $\hat\rho^N$. This expression contains a product of $N$ vacuum matrix elements of operators that depend on the field $\phi$. Each one of these matrix elements is calculated independently, and thus the fields entering the calculation of different matrix elements can be considered as independent. We thus define the multiplet of {\it replica fields} $\phi_i^\alpha$, $\alpha=1,...,N$.
Thus we can write
 \beq
tr[\hat\rho^N ]=  \langle 0\vert e^{ -\sum_{\alpha=1}^N \phi_i^\alpha M_{ij}\phi_j^\alpha \,+\,\sum_{\alpha=1}^N\phi_i^\alpha M_{ij}\phi_j^{\alpha+1}} \, 
\vert 0\rangle
\label{rhoN}
\eeq
where now $\vert 0\rangle$ is the light front vacuum of all the replica fields $\phi^\alpha$.
Notice the nearest neighbor "interaction"  between the replica fields. 
The replica fields in eq.(\ref{rhoN}) satisfy periodic boundary conditions $\phi^{N+1}=\phi^1$. 

 We further rewrite (\ref{rhoN})  
   \beq
tr[\hat\rho^N ]=  \left(\frac{\det[\pi]}{2\pi}\right)^{N/2} \int \prod_{\alpha=1}^N[D\phi^\alpha]\,
 \exp\left\{ -\frac{\pi}{2}\sum_{\alpha=1}^N \phi_i^\alpha \phi_i^\alpha \,-\,\frac{1}{ 2}\sum_{\alpha=1}^N
(\phi_i^\alpha - \phi_i^{\alpha+1})\,M_{ij}\,(\phi_j^\alpha - \phi_j^{\alpha+1}) \right\} \, 
\label{rhoN1}
\eeq
This is a partition function of a spin chain  on a replica space lattice.
 Eq.(\ref{rhoN1}) obviously has the discrete translational symmetry in the replica space.
The "action" is diagonalized by Fourier transforming in $\alpha$:
\beq\label{FT}
\tilde \phi^n\,=\,\frac{1}{N}\,\sum_{\alpha=1}^Ne^{i\frac{2\pi}{ N}\alpha n}\ \phi^\alpha\,; \ \ \ \ \ \ \ \ \ \ \ \ \ \  \phi^\alpha\,=\,\sum_{n=0}^{N-1}e^{-i\frac{2\pi}{ N}\alpha n}\ \tilde\phi^n
\eeq
Notice the periodicity relation $\tilde\phi^{N-n}=\tilde\phi^{*n}$.
The nearest-neighbor  interaction in Fourier space reads
\beq
(\phi_i^\alpha - \phi_i^{\alpha+1})(\phi_j^\alpha - \phi_j^{\alpha+1})= \sum_{n,m}\left(e^{-i\frac{2\pi}{N} n}-1\right)\left(e^{-\frac{2\pi}{N} m}-1\right)
e^{-i\frac{2\pi}{N}\alpha (n+m)}\,\tilde\phi^n_i\tilde\phi^m_j
\eeq 
 Using 
 \beq
 \sum_{\alpha}e^{-i\frac{2\pi }{ N}\alpha (n+m)}\,=\,N\,\delta_{n+m,N}
 \eeq 
 we have
 \beq
\sum_{\alpha}(\phi_i^\alpha - \phi_i^{\alpha+1})(\phi_j^\alpha - \phi_j^{\alpha+1})= N\,\sum_{n}\left(e^{-i\frac{2\pi}{ N} n}-1\right)\left(e^{-i\frac{2\pi}{ N} n}-1\right)\,
\tilde\phi^n_i\tilde\phi^{*n}_j\,=\,4\,N\,\sum_{n}\,\sin^2(\frac{\pi}{ N}\,n)\,\tilde\phi^n_i\tilde\phi^{*n}_j
\eeq 
   \beq
tr[\hat\rho^N ]= N^{N/2}\, \left(\frac{\det[\pi]}{ 2\pi}\right)^{N/2}  \,\int \prod_{n}[D\tilde \phi^n]\,
 \exp\left\{ -\frac{N}{2}\sum_{n=0}^{N-1} \tilde\phi_i^n\, \left[\pi\,+\, 4\,M\,\sin^2(\frac{\pi}{ N}\,n)\,\right]_{ij}\, \tilde\phi_j^{*n} \right\} \, 
\label{rhoN3}
\eeq
where $N^{N/2}$ is the Jacobian of the transformation (\ref{FT}). Using the identity 
\beq
\pi\,+\, 4\,M\,\sin^2(\frac{\pi}{ N}\,n)\,=\,\left[ \pi\,+\,2 M\left(1-\,\cos ( \frac{2\pi}{ N}\,n)\right) \right]\,
\eeq
the Gaussian integral in (\ref{rhoN3}) is easily computable
\beq
tr[\hat\rho^N ]=  \det[\pi]^{N/2}\,\det\left[\prod_{n=0}^{N-1} \left[ \pi\,+\,2 M\left(1-\,\cos ( \frac{2\pi}{ N}\,n)\right) \right]^{-1/2}\right]
\eeq
Now we apply (1.396) of Ref. \cite{GR}, with $1>2MG>0$:
\bea 
\prod_{n=0}^{N-1}\left[ \pi\,+\,2 M\left(1-\,\cos ( \frac{2\pi}{ N}\,n)\right) \right]&=&(2 M)^N\, \prod_{n=0}^{N-1} \left[\,1-\,\cos ( \frac{2\pi}{ N}\,n) +\frac{\pi}{2M}\right]  \nn\\
&=&\,2\,(M)^N\,\left[\cosh (N\,\arccosh[1\,+\,\frac{\pi}{2M}])\,-\,1\right]
\eea
We finally arrive at the following expression for $tr[\hat\rho^N]$:
\beq
tr[\hat\rho^N ]\,= \,\exp \left\{ -\,\frac{1}{ 2}\ln 2 \,-\,\frac{N}{ 2}\,tr\Big[\ln \frac{M}{\pi}\Big]\,-\,\frac{1}{2}\,
tr\Big[\ln\Big(\cosh (N\,\arccosh[1+\frac{\pi}{2M}])-1\Big)\Big]\right\}
\label{rhoN5}
\eeq
The entropy is computed by taking $N=1+\epsilon$ and keeping the terms linear in $\epsilon$:
\bea
tr[\hat\rho^{1+\epsilon}]&\simeq&\exp \left\{ -\frac{1}{ 2}\ln 2 -\frac{1+\epsilon}{ 2} tr\left[\ln \frac{M}{\pi}\right]-\frac{1}{2}
tr\left[\ln\left( \frac{\pi}{2M}+\epsilon  \frac{\pi}{2M}\sqrt{1+\frac{4M}{\pi}} \,\arccosh[1+\frac{\pi}{2M}] \right)\right ]\right\} \nn \\ && \nn\\
&=&1\,+\,\frac{\epsilon}{ 2}\,tr\left[\ln \frac{\pi}{M}  \,-\,\sqrt{1+\frac{4M}{\pi}}\arccosh[ 1+\frac{\pi}{2M}]      \right]
\eea
Thus we arrive at a closed expression for the entanglement entropy
\begin{eqnarray}\label{sigma}
\sigma^E=\frac{1}{2}\,tr\left[\ln \frac{M}{\pi}  \,+\,\sqrt{1+\frac{4M}{\pi}}\arccosh[ 1+\frac{\pi}{2M}]  \right] =\frac{1}{2} tr\left\{\ln \frac{M}{\pi}+ \sqrt{1+\frac{4M}{\pi}}\ln\left[1+\frac{\pi}{2M}\left(1+\sqrt{1+\frac{4M}{\pi}}\right)\right]  \right\}\nn \\
\end{eqnarray}


To understand the basic properties of this expression let us consider translationally invariant case. In this case the matrix $M$ is diagonal in momentum space,
\beq
M^{ab}_{ij}(p)\,=\,g^2\,\mu^2(p^2)\,\frac{p_ip_j}{p^4}\,\delta^{ab}
\eeq
In the original MV model $\mu^2$ is a constant and does not depend on momentum, but the momentum dependent Gaussian width has been used in recent applications (see e.g. \cite{DL1}). The contribution to the entropy from large transverse momentum modes can be calculated by expanding the expression eq.(\ref{sigma}) to leading order in $M$, since for large momenta ($g^2\mu^2<p^2$) the eigenvalues of $M$ are all small.
The expression for the entropy in the weak field limit is
\begin{equation}\label{weak}
\sigma^E_{M\ll 1}\,=\,tr \left[\frac{M}{\pi}\ln\frac{\pi e}{M}\right]
\end{equation}
Here we have kept the first two terms in the small $M$ expansion, which are not suppressed by powers of $M$.

Thus the dominant ultraviolet contribution  is
\beq\label{weak1}
\sigma^E_{UV}
\,\simeq\, -\,\frac{g^2}{\pi}\,(N_c^2-1)\, S\,\int \frac{d^2p}{(2\pi)^2}\, \frac{\mu^2(p^2)}{p^2}\,\ln \frac{g^2\mu^2(p^2)}{e\pi\,p^2}\,
\theta(\,p^2\,-\,\frac{g^2}{\pi}\,\mu^2(p^2))
\eeq
where $S$ is the total area of the projectile. 
For the original MV model with momentum independent $\mu$ this expression is logarithmically divergent in the UV. Introducing
a UV cutoff $\Lambda$ we find 
\beq\label{UV}
\sigma^E_{UV}\,\simeq\,\frac{Q_s^2}{4\pi g^2}(N_c^2-1)S\left[\ln^2\frac{g^2\Lambda^2}{Q_s^2}+\ln\frac{g^2\Lambda^2}{Q_s^2}\right]
\eeq
where we have identified the saturation momentum in the standard way as $Q_s^2=g^4\,\mu^2/\pi$.
 The UV divergence in this expression is of course the artifact of the eikonal approximation which in the CGC context is applied for all transverse momentum modes. In fact the eikonal approximation breaks down when the transverse momentum of the soft gluons is of the order of their longitudinal momentum. The natural cutoff on the transverse momentum is of order $\Lambda\sim P^+\,e^{-Y}\sim me^{Y_0}$, where $P^+$ is the total energy of the hadron and $m$ is a typical hadronic scale and $Y_0$ is the rapidity of the soft gluons. This cutoff does not depend on the total energy of the hadron, but rather on some initial energy at which the eikonal approximation is assume to be valid\footnote{ $\sigma^E_{UV}$ given by (\ref{weak1}) superficially bears some similarity to the expression for dynamical entropy proposed in 
\cite{Peschanski}.  The two ``entropies'' are however defined differently and evaluate to different values.}.

The contribution from the infrared modes can also be calculated. At small momenta $p^2<Q_s^2/g^2$ one can formally expand in $M^{-1}$. In this limit the entropy is dominated by the first term in eq.(\ref{sigma}). Here we also keep the first subleading correction:
\beq\label{strong}
\sigma^E_{M\rightarrow\infty}\,\simeq\,\frac{1}{2}\,tr [\ln\frac{e^2M}{\pi} ]  
\eeq
and the infrared contribution is 
\beq\label{IR}
\sigma^E_{IR}\, \simeq\,\frac{1}{2}\,(N_c^2-1) \,S\, \int \frac{d^2p}{(2\pi)^2}
\,\ln \frac{e^2\,g^2\,\mu^2(p^2)}{\pi\,p^2}\, \theta(Q_s^2\,-\,g^2p^2)=\frac{3}{8\pi g^2}(N_c^2-1)SQ_s^2
\eeq
Note that we have chosen to separate the integration region into UV and IR at exactly $p^2=Q_s^2/g^2$ in eqs.(\ref{weak1},\ref{IR}). Although the parametric dependence of the separation scale is clear, its exact value is somewhat arbitrary. Our reason for choosing the above value, is that at this value of $p^2$ the integrands in eqs.(\ref{weak1},\ref{IR}) exactly coincide. Thus this is a unique choice for which the approximation of the total momentum space entropy density by the sum of its asymptotic expressions is a continuous function of momentum. 

Combining the two expressions we find approximately for the MV model
\beq\label{sigmaf}
\sigma^E\approx\sigma^E_{UV}+\sigma^E_{IR}=\frac{SQ_s^2}{4\pi g^2}(N_c^2-1)\left[\ln^2\frac{g^2\Lambda^2}{Q_s^2}+\ln\frac{g^2\Lambda^2}{Q_s^2}+\frac{3}{2}\right]
\eeq


A word on the dependence of the entropy on the energy (rapidity) of the hadron.
The above calculation is performed at fixed rapidity. The hadronic wave function evolves to higher rapidity via JIMWLK evolution. Since this evolution is nonlinear, the soft gluon wave function is not given by a Gaussian any more, especially in the saturation regime.
Calculating the entropy for a nongaussian wave function is significantly more difficult, although we will describe some steps toward such a calculation in the last section. However, as has been suggested in the past,  approximating the soft gluon wave function by a gaussian is not a bad approximation for phenomenological purposes as long as the parameters of the Gaussian are taken to evolve with rapidity in a way consistent with JIMWLK evolution\cite{IIM}.

In particular one can still use the MV type ansatz but with the function $\mu_Y(p^2)$ taken to be a solution of the  BFKL\cite{bfkl} (or BK\cite{Bal,bk}) evolution equation. 
Within this approximation one can derive the evolution of the entropy with the total energy of the hadron. In the weak coupling limit, where the evolution of $\mu$ is given by the BFKL equation, differentiating the weak field expression eq.(\ref{weak}) we find
\beq\label{evol}
\frac{d\sigma^E}{dY}\,=\,- (N_c^2-1)S\int\frac{d^2p}{(2\pi)^2}\frac{d^2k}{(2\pi)^2}\ln [\phi(p^2)]K_{BFKL}(p,k)\phi(k^2)
\eeq 
where $\phi(k^2)\equiv \frac{g^2}{\pi}\frac{\mu^2(k^2)}{k^2}$ is the gluon unintegrated density, and $K_{BFKL}$ is the kernel of the BFKL equation.

In contrast to the MV model, the solution of the BFKL/BK equation at large rapidity exhibits an anomalous dimension which slowly varies with transverse momentum
\beq \frac{g^2}{\pi}\mu^2_Y(p^2)\sim p^{\gamma(p)};
\eeq
with $\gamma(p>>Q_s)\rightarrow 1$;  $\gamma(p\sim Q_s)\approx .67$ and $\gamma(p<Q_s)=0$.  
Note that with such an anomalous dimension the entropy diverges as a power of the cutoff in the ultraviolet as $\sigma_{UV}\propto\Lambda^\gamma$.

Some properties of the entanglement entropy of CGC are worth noting. First, it is proportional to the transverse area of the hadron. This is a natural property of any extensive observable. However $\sigma^E$ is not extensive in the longitudinal direction. Recall that $\sigma^E$ as calculated above is the entropy on the rapidity slice of width $\Delta Y$, but $\sigma^E$ is not proportional to $\Delta Y$. The reason for this is of course clear, since only a single rapidity independent mode on the interval $\Delta Y$ is entangled with the valence degrees of freedom of the hadron. For the same reason the entropy does not have any longitudinal UV divergence associated with the fact that we have a sharp boundary between soft and valence degrees of freedom in rapidity space. This is very different from a situation one encounters in calculating entanglement entropy of a finite region of space $A$ in a local field theory when integrating out degrees of freedom in the rest of space $B$. In the latter case the local interaction between the degrees of freedom along the boundary between $A$ and $B$ leads to a UV divergence which depends on the ratio of the correlation length to the (vanishing) width of the boundary region. In our case the interaction in the longitudinal direction is nonlocal. The eikonal interaction between soft and valence degrees of freedom extends over large distances in rapidity, so that there is no significant contribution to the entropy from the (sharp) boundary. In this sense the entanglement entropy of the CGC is akin to the ``topological entropy''\cite{topological} frequently discussed in the context of condensed matter systems, which is a constant independent of the boundary or volume of the spatial region under consideration.

Recently, the ``momentum space'' entanglement entropy has been discussed in connection with Wilsonian renormalization group flow\cite{momentum entropy}. It would be interesting to explore the relation of these ideas with eq.(\ref{evol}) as $\sigma^E$ is indeed calculated by separating degrees of freedom in momentum space, while the BFKL equation has been interpreted in the past as RG flow in rapidity variable\cite{jkmw}.

\section{Entropy production in collision.}

Our next quantity of interest is the entropy production during collision of two hadrons.
We calculate the entropy in the rapidity bin of width $\Delta Y$ at rapidity $Y$ away from the valence charges. All the glue at rapidities $\eta<Y+\Delta Y$ relative to the forward moving particles is considered to be part of the target wave function and as in the standard saturation approach is represented by the eikonal scattering matrix $S(x)$.


According to the eikonal paradigm the effect of the collision is to color rotate the valence as well as the soft gluon field in the wave function, so that at time $t=0$ right after the collision

\beq
\psi[A]\chi[a,\rho]\rightarrow \hat S\psi[A]\chi[a,\rho]=\psi[SA]\chi[Sa,S\rho]=\psi[SA]\exp\left\{i\int _x
 \tilde b^i_a(x)S(x)\left[a^{\dagger i}_a(x)+a^i_a(x)\right]\right\}\vert 0\rangle,
\eeq
where $\tilde b^a_i(x)$ is the Weszacker-Williams field produced by the rotated color charge density $\tilde\rho^b(z)=S^{bc}(z)\rho^c(z)$. Here $\hat S$ is the second quantized operator of eikonal $S$-matrix. 
 The action of $\hat S$ to the right (and $\hat S^\dagger$ to the left) is equivalent to the multiplication of $a^a(x),a^{\dagger a}(x)$ and $\rho^a(x)$ by the unitary matrix $S(x)$. 
Since $\hat S$ is a unitary operator which does not mix valence and soft degrees of freedom, it is clear that the initial scattering does not modify the entanglement entropy. 
To see this explicitly we note that in the basis  $\tilde\rho(x) =S(x)\rho(x)$ and $\tilde\phi(x)= S(x)\phi(x)$ the form of the scattered wave function is identical to that of the incoming one in the original basis. Since the integration measure for $\rho$ as well as
the vacuum wave function for $\phi(x)$ eq. (\ref{vac}) are invariant under this local unitary transformation, changing variables to $\tilde\rho$ and $\tilde \phi$ in the calculation of the entropy immediately establishes that the entropy is the same.

However the entropy produced in collision is not the same as the entanglement entropy of the final state.
Here one is interested only in characteristics of inelastically produced gluonic state, and not, say in the contribution of soft gluons which are part of the wave function of outgoing ``bound state'' hadrons.
Thus the relevant quantity for calculating the production entropy is not the wave function evolved to $t\rightarrow\infty$, but only its ``inelastically produced'' part.
This is a similar problem to the one we are faced with when calculating any soft gluon observable produced in collision, i.e. single or double gluon inclusive cross section. In that context the solution is well known. Recall, that a gluonic observable $O(a, a^\dagger)$ at $t\rightarrow \infty$ is given by the  expression \cite{KLincl}
\begin{equation}
 \langle O\rangle\,=\,\int d\rho W[\rho]\langle 0|e^{-i\int _x
 b^i_a[\rho](x)\phi^i_a(x)}\hat S^\dagger e^{i\int _x
 b^i_a[\rho](x)\phi^i_a(x)}O(a,a^\dagger)e^{-i\int _x
 b^i_a[\rho](x)\phi^i_a(x)}\hat Se^{i\int _x
 b^i_a[\rho](x)\phi^i_a(x)}|0\rangle\end{equation}
The multiplication by the extra coherent operator factor accounts for the evolution of the wave function from time $t=0$ to time $t\rightarrow \infty$ as well as restricting Hilbert space for the calculation of $O$ to that of inelastically produced gluons only. 
 
 Since this formula is valid for an arbitrary soft observable $O(a,a^\dagger)$, clearly this  means  that the density matrix that describes the {\it produced} soft gluons has exactly the same form as eq.(\ref{rho1}) but with the matrix $M$ substituted by
\beq\label{Minf}
\,M^P\equiv \,\frac{g^2}{ 4\pi^2} \int_{u,v} \mu^2(u,v) \frac{(x-u)_i}{(x-u)^2}\frac{(y-v)_j}{(y-v)^2}\,[(S(u)-S(x))(S^\dagger(v)-S^\dagger(y))]^{ab}
\eeq
Thus the entropy produced in a hadronic collision, which is the Van Neumann entropy of the density matrix specified by eq.(\ref{Minf}) is given by 
\beq\label{sigmaprod}
\sigma^P =\frac{1}{2} tr\left\{\ln \frac{M^P}{\pi}+ \sqrt{1+\frac{4M^P}{\pi}}\ln\left[1+\frac{\pi}{2M^P}\left(1+\sqrt{1+\frac{4M^P}{\pi}}\right)\right]  \right\}
\eeq
Although the formal expression for $\sigma^P$ is similar to that for $\sigma^E$, it has a different meaning and its properties are significantly different.
We stress that $\sigma^P$ is not an entanglement entropy like $\sigma^E$, but rather the entropy of the state produced inelastically in the hadronic collision. As a result $\sigma^P$ is UV finite. This is clear, since the UV divergence in the calculation of $\sigma^P$ came from the coordinate region $x\rightarrow y\rightarrow u\rightarrow v$, and $M^P$ vanishes in this limit, as opposed to $M$.

Another important property is that $\sigma^P$ as defined in eq.(\ref{sigmaprod}) is calculated for a single scattering event characterized by a given eikonal profile $S(x)$. Although it certainly makes sense to talk about event by event entropy production, a more global characteristic of scattering would be an event averaged quantity. We will next consider such an average production entropy and will relate it to inclusive gluon production amplitudes.

Note, that eq. (\ref{Minf})  can be written as a product of single inclusive gluon production amplitude $Q_i$ defined in \cite{KLincl} averaged over the projectile wave function
\beq
\,M^P_{ij}\,=\, \int D[\rho]\, Q_i\,Q_j\,W_P[\rho]\,\equiv\,\langle Q_i\,Q_j\rangle_P
\eeq
where
\beq
Q_i^a(x)\,=\,\frac{g}{2\pi}\,\int_{u} \frac{(x-u)_i}{(x-u)^2}\,[S(u)-S(x)]^{ab} \rho^b(u)
\eeq
with $S$ being in  the adjoint representation.

The target average of this expression is directly related to the single inclusive gluon production probability, or rather the phase space density $n=dN/dkdydb$ is
\beq
n(k)=\frac{dN}{d^2kdyd^2b}\,=\, \int_{xy}\langle \langle Q_i^a(x)\,Q_i^a(y)\rangle\rangle_{P,T}\,e^{ik(x-y)}
\eeq
where as usual  the averaging over the target fields has to be performed with some weight function $W[S]$.

In order to relate the average entropy to inclusive gluon production amplitudes we expand it in powers of the fluctuation of the matrix $M^P$ around its target average 

\beq
M^P= \bar M^P \,+\, (M^P\,-\,\bar M^P);\ \ \ \ \ \ \ \ \ \ \ \ \ \  \ \ \ \bar M^P\,\equiv\, \int DS\,W[S]\, M^P[S]
\eeq

The target average has the form
\beq\label{pomerons}
\bar M^P\,=-\,\delta^{ab} \,\frac{g^2}{ 4\pi^2}\, \int_{u,v} \mu^2(u,v) \frac{(x-u)_i}{(x-u)^2}\frac{(y-v)_j}{(y-v)^2}\,[P_A(x,y)+P_A(u,v)-P_A(x,v)-P_A(u,y)]
\eeq
where the ``adjoint Pomeron'' is defined as
\beq P_A(x,y)=1-\frac{1}{N_c^2-1}\langle tr [S^\dagger(x)S(y)]\rangle_T\eeq
Although $\bar M^P$ is related to the single inclusive gluon production probability, the two are not equal to each other.


Assuming translational invariance of $\mu$ and $P_A$ we have
\beq\label{pomerons1}
\bar M^P\,=\,\,\delta^{ab} \,g^2\,\int_{k,q} \,e^{-ik(x-y)}\,K_{ij}(k,q) \,\mu^2(k-q)\,P_A(q)
\eeq
where
\beq
K_{ij}(p,q)= \frac{(k-q)_i}{ (k-q)^2}\frac{(k-q)_j}{ (k-q)^2}\,+\, \frac{k_i}{ k^2}\frac{k_j}{k^2}\,-\, \frac{(k-q)_i}{ (k-q)^2}\frac{k_j}{ k^2}\,-\,
\frac{(k-q)_j}{ (k-q)^2}\frac{k_i}{k^2}
\eeq
Conveniently assuming parity invariant target, we can decompose $\langle M^P\rangle_T$ in momentum space as
\beq\label{Mproj}
\bar M^P(k)\,=\,\delta^{ab}\,\left[M_l(k^2)\,\frac{k_ik_j}{k^2}\,+\,M_t(k^2)\,(\delta_{ij}\,-\,\frac{k_i\,k_j}{k^2})\right]
\eeq
As opposed to $M$, the matrix $M^P$ has both longitudinal and transverse components. These correspond to gluons produced with polarizations parallel and perpendicular to their transverse momentum. Clearly both polarizations in general are produced, since the gluons in question are free gluons rather than a part of the dressing of the wavefunction of some faster moving valence color charges. The inclusive gluon production cross section 
\cite{KT} is related to the sum of the two polarizations
\beq n(k)\propto M_l(k)+M_t(k)\eeq

%

Let us consider the case of small $M^P$, which is relevant for the UV regime. We can then use eq.(\ref{weak})
\beq\label{Expand}
\bar \sigma^P= \langle tr \left[\frac{M^P}{\pi}\ln\frac{\pi e}{M^P}\right]\rangle_T= tr \left[\frac{\bar M^P}{\pi}\ln\frac{\pi e}{\bar M^P}\right]
 -\frac{1}{2\pi}\,
 tr \left [ \Big\{\langle M^P\,  \,M^P\rangle_T\,- (\bar M^P)^2\Big\} (\bar M^P)^{-1}\right ] ....
\eeq
Using the representation (\ref{Mproj}), the first term in (\ref{Expand}) can be written as
\beq\label{leading}
tr \left[\frac{\bar M^P}{\pi}\ln\frac{\pi e}{\bar M^P}\right]=-\frac{N_c^2-1}{\pi}\int \frac{d^2k}{(2\pi)^2}\Big[M_l(k^2)\ln \frac{M_l(k^2)}{e\pi}+M_t(k^2)\ln \frac{M_t(k^2)}{e\pi}\Big]
\eeq 
This has the standard form $\sigma=-\sum_in_i\ln n_i$ for the system of noninteracting particle species. The index $i$ here refers to transverse momentum as well as the longitudinal and transverse gluon polarizations. Note, that it cannot be expressed in terms of the total particle spectrum $n(k)$, unless  the longitudinal and transverse eigenvalues of $M^P$ are proportional to each other. Generically we do not see a reason to expect that this is the case.


The second term in (\ref{Expand})  is related to correlated inclusive two gluon production \cite{jamal,Braunincl2,DMV,ddgjlv,adrianjamal,DV,correlations,LR,DLM}. Pictorially this term can be represented in Fig.1a.  This diagram (apart from the fact that two of the gluon polarization indices do not close) is the same as the connected ``glasma graph'' contribution to the two gluon correlation \cite{DMV}. The ``disconnected'' diagram in Fig.1b is not included in the expression eq.(\ref{Expand}). Note that although the diagram in Fig. 1b has a disconnected topology, it does in fact contribute to the correlated production in the large $N_c$ limit if the target averages do not factorize\cite{correlations, dumitru-skokov}. In the language of the Reggeon field theory, this contribution is due to correlated production from two Pomerons, while the diagrams in Fig 1a are due to correlated production from the B-reggeon \cite{Breggeon}. 

 \begin{figure*}[htp]
\vspace*{-0.2cm}
\includegraphics[width=12cm]{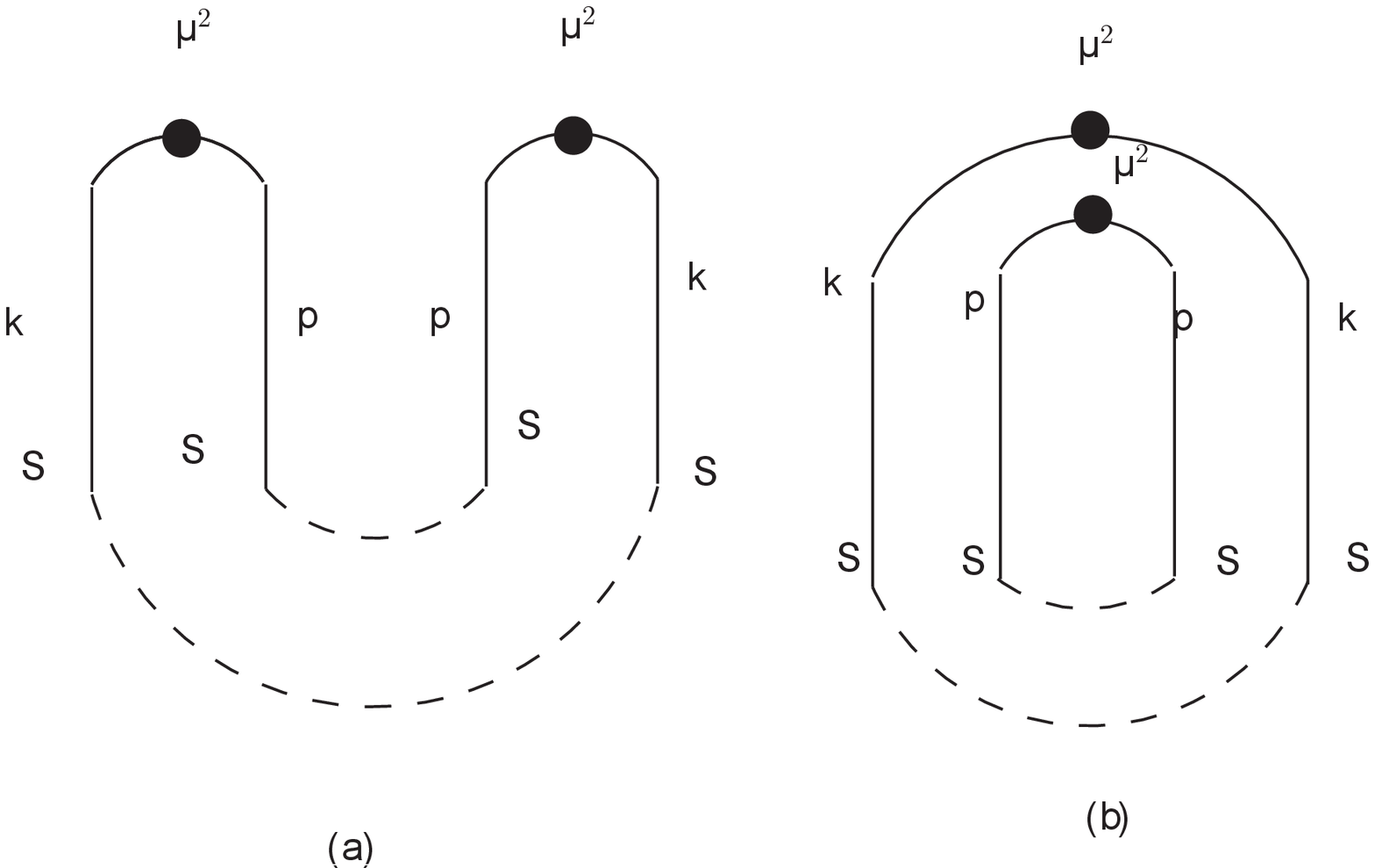}
{\caption  \ Two contributions to the inclusive two gluon production. Fig.1a - ``connected'' part that contributes to entropy eq.(\ref{Expand}). Fig. 1b - ``disconnected part that does not contribute to the entropy eq.(\ref{Expand}). }
\end{figure*}

With this in mind we can write 
\begin{eqnarray} \label{entcorr}
&&\delta\sigma^P=\frac{1}{2\pi}\,
 tr \left [ \Big\{\langle M^P\,  \,M^P\rangle_T\,- (\bar M^P)^2\Big\} (\bar M^P)^{-1}\right ]\\
 &&\propto
 \int d^2kd^2p\Big[\langle a^{\dagger a}_l(p)a^{\dagger b}_i(k)a^c_j(k)a^a_l(p)\rangle_{(B)}-\langle a^{\dagger a}_l(p)a^a_l(p)\rangle\langle a^{\dagger b}_i(k)a^c_j(k)\rangle\Big](\bar M^{P})^{-1\,cb}_{\ \ \ ji}(k)\nonumber\end{eqnarray}
 Or in a less convoluted way:
 \begin{eqnarray}
\delta\sigma^P&= &- \frac{g^4}{ 32\,\pi^5\,(N_c^2-1)} \int_{u,v,\bar u,\bar v, y,x,\bar x,k} \mu^2(u,v) \mu^2(\bar u,\bar v)\,e^{ik(x-\bar x)} \left[\frac{1}{M_l(k^2)}\,\frac{k_ik_j}{k^2}\,+\,\frac{1}{ M_t(k^2)}\,(\delta_{ij}\,-\,\frac{k_i\,k_j}{k^2})\right]\nonumber \\
&&\times \frac{(x-u)_i}{(x-u)^2}\frac{(y-v)_m}{(y-v)^2}\, \frac{(\bar x-\bar u)_j}{(\bar x-\bar u)^2}\frac{(y-\bar v)_m}{ (y-\bar v)^2}\nonumber \\
&&\times\, tr \langle \left[(S(u)-S(x))(S^\dagger(v)-S^\dagger(y))][(S(\bar u)-S(\bar x))(S^\dagger(\bar v)-S^\dagger( y))\right]\rangle_T\,+\,
\frac{1}{2\,\pi}\,\int_k n(k)
\end{eqnarray}

 Again, we see that if the longitudinal and transverse parts of $M$ are proportional to each other at all momenta, the extra contribution to entropy is proportional to the double inclusive gluon correlation function integrated over the gluon momenta. If $M_l(k)\ne aM_t(k)$, the longitudinal and transverse gluon polarizations are separately normalized to their single inclusive production probability.
 
Interestingly the correlation function in eq.(\ref{entcorr}) is normalized by dividing by the first power of the single inclusive gluon production and not its square. Recall that the correlated gluon production normalized this way remains finite in the CGC approach in the limit $Q_s^2S\rightarrow\infty$, while the correlation function normalized to the square of single inclusive gluon vanishes in this limit\cite{DV, correlations}.


The natural, but nevertheless interesting property of eq.(\ref{Expand}), is that the contribution of correlations to entropy is negative. This is in accordance with the intuition that a correlated state contains more order and thus has a smaller entropy.  

As noted above, the produced entropy is UV finite, since in the UV limit all adjoint Pomerons in eq.(\ref{pomerons}) vanish, and the divergent contribution to the momentum integral in eq.(\ref{Expand}) disappears. 
This is even clearer from the momentum space representation of $M^P$ eq.(\ref{pomerons1}). The perturbative behavior of the adjoint Pomeron at large momenta is $P(q)\propto Q_T^2/q^4$, which leads to $M^P(k\rightarrow\infty)\rightarrow Q_p^2Q_T^2/g^2k^4$. Even accounting for the BFKL anomalous dimension leads to the UV convergent integral in eq.(\ref{leading}).
The integral in eq.(\ref{leading}) is also IR finite, since the expected IR behavior is $P(q\rightarrow 0)\rightarrow 1/Q_T^2$. 

In general the consequence of eq.(\ref{pomerons}) (neglecting the rotational indices for the moment) is
\beq \label{mpar}M^P(q)=\pi \frac{Q_P^2}{g^2Q_T^2} F\left(\frac{Q_T^2}{q^2}\right)\eeq
where $F$ is a regular function with the limiting behavior $F(x\rightarrow 0)\rightarrow const\, x^2$ and $F(x\rightarrow\infty)\rightarrow const$.
Using eq.(\ref{mpar}) we can estimate the entropy production parametrically

\beq \label{sest}
\sigma^P\sim a(N_c^2-1)S\frac{Q_P^2}{\pi g^2}\ln \frac{b g^2Q_T^2}{Q_P^2}
\eeq
where $a$ and $b$ are constants of order unity.

\section{Discussion.}
In this paper we have calculated the entanglement entropy of the soft gluon modes in the CGC wave function, as well as entropy produced in hadronic collision at high energy. In these calculations we have relied on the MV model to represent the valence part of the hadronic wave function. 

The entanglement entropy is an interesting theoretical quantity, as it is a global characteristic of the soft gluon density matrix. 
Its basic properties are determined by the fact that only one longitudinal mode of the soft gluon field, namely the rapidity independent mode is entangled with the valence part of the wave function. The entropy  therefore is not extensive in the length of the rapidity interval, and in this sense is akin to the topological entropy discussed these days in condensed matter literature\cite{topological}. In a certain sense the analogy is closer than it might seem, as the origin of the topological entropy is in the long range quantum entanglement.

We find that the entanglement entropy of CGC is formally UV divergent and is dominated by the modes with large transverse momentum 
$p_\perp^2>Q_s^2/g^2$. The contribution from the IR modes is finite and suppressed by a square of the logarithm of the UV cutoff.

We have also derived an expression for the entropy production in an hadronic collision. We have shown that it is related to inclusive gluon production amplitudes, but is not determined exclusively in terms of these amplitudes. In particularly it depends on the production probabilities of longitudinally and transversely (with respect to the direction of their transverse momentum) polarized gluons separately. We also find that correlated gluon emission gives a negative contribution to the produced entropy, consistent with the view of entropy as measuring disorder in the final state. Just like the entanglement entropy, the produced entropy is not extensive in the rapidity interval over which it is calculated. The reason is the same: in the CGC approximation  all final state particles are produced from a single longitudinal boost invariant mode.

Having calculated the entropy, the natural question is whether one can define the corresponding temperature. The natural definition would be the usual thermodynamic one
\beq T^{-1}=\frac{d\sigma}{d E_\perp}\eeq 
where $ E_\perp$ is the transverse energy of the system.

It is not clear whether this is a sensible quantity for the entanglement entropy. If for $E_\perp$ we take the obvious 
\beq E_\perp\propto\int d^2k|k|M(k)\propto \frac{Q_s^2 S}{g^2}\Lambda\eeq
we obtain for the ``entanglement temperature''
\beq T_E\propto \frac{\Lambda}{(N_c^2-1)\ln \frac{\Lambda^2 g^2}{Q_s^2}}\rightarrow \infty
\eeq
 which reflects the fact that the entanglement entropy is dominated by the UV modes and is formally UV divergent.

 On the other hand for the produced system of particles eq.(\ref{pomerons1}) gives
 \beq E_\perp\propto (N_c^2-1)S\frac{Q_P^2}{g^2}Q_T\eeq
This is just the statement that the average transverse momentum of produced particles is proportional to $Q_T$, while the number of particles produced is of the order $Q_P^2/g^2$.

Since the entropy eq.(\ref{sest}) is not only a function of $E_\perp$, the definition of temperature is somewhat ambiguous. To define the temperature one has to decide which  combination of $Q_P$ and $Q_T$ has to be kept fixed. Choosing which combination to keep fixed corresponds to deciding what is the analog of increasing the energy in the {\it same} statistical mechanical system.  One reasonable choice is to keep $Q_P$ fixed, since this corresponds to increasing the energy of the produced system by boosting the target, while keeping the projectile unchanged. With this choice using eq.(\ref{sest}) we find
\beq\label{t}
T=c\, Q_T=\frac{\pi}{2}\langle k_\perp\rangle
\eeq
where $c$ is a constant of order unity and $\langle k_\perp\rangle=E_\perp/N_{total}=\frac{\int d^2k |k|M(k)}{\int d^2pM(p)}$ is the average transverse momentum per produced particle. Note, that even though the relation between $\langle k_\perp\rangle$ and $Q_T$ depends on the details of the function $F$ in eq.(\ref{mpar}), the relation between $\langle k_\perp\rangle$ and temperature is determined exactly. Eq.(\ref{t}) is an intuitively simple result showing that the effective temperature is proportional to average momentum per particle. We are nevertheless not aware of a direct derivation of this result from the density matrix of the system produced in collision.

All our calculations were performed in the MV model. The Gaussian weight for the valence charge density averaging made it possible to express the entropy directly in terms of a single function $M^P$. The actual weight functional evolves with energy and is determined by solving the JIMWLK equation. Although the calculation of entropy in the case of non Gaussian $W[\rho]$ is more complicated, one can try to approach it in a similar way.

We would have to calculate
\begin{equation} \label{nonG} tr[\hat \rho^N]=tr\int \prod_{\beta=1}^N D[\rho_\beta] \,W_P[\rho_\beta]\
e^{i\int_k b_\beta(k)\phi(k)}\vert 0\rangle\langle 0\vert e^{-i\int_k b_\beta(k)\phi(k)}
\end{equation}

Integrating over $\rho_\beta$, and resorting to the same replica trick yields
\begin{eqnarray} tr[\hat \rho^N]&=& \left(\frac{\det[\pi]}{ 2\pi}\right)^{N/2} \int \prod_{\alpha=1}^N[D\phi^\alpha]\,
 \exp\left\{ -\frac{\pi}{2}\sum_{\alpha=1}^N \phi_i^\alpha \phi_i^\alpha \,+\,\frac{1}{ 2}\sum_{\alpha=1}^N \Gamma[\phi^\alpha-\phi^{\alpha+1}] \right\} \, 
\label{rhoW}
\end{eqnarray}
Here as before we have introduced $N$ replicas - one for each vacuum matrix elements.
In eq.(\ref{rhoW}) $\Gamma$ is the effective action.
Interestingly, it is directly related to the projectile scattering matrix, since
\begin{equation}
S=\int D\rho W_P[\rho]e^{ig\int_{x,u}\rho^a(x)\frac{(x-u)_i}{(x-u)^2}[\phi_i^\alpha(u)-\phi_i^{\alpha+1}(u)]}\equiv e^{-\frac{1}{2}\Gamma[\phi^\alpha-\phi^{\alpha+1}]}
\end{equation}
is precisely the scattering matrix element of the projectile on the ``target field'' $\phi_i^\alpha(u)-\phi_i^{\alpha+1}(u)$.

The action $\Gamma$ can be expanded in terms of connected correlators of the valence color charge density
\beq
\Gamma[\phi]\,=\,\Gamma^2_{ij} \phi_i\phi_j\,+\, \Gamma^3_{ijn}\phi_i\phi_j\phi_n\,+\,\Gamma^4_{ijnm}\phi_i\phi_j\phi_n\phi_m\,+\,...
\eeq
\begin{eqnarray}
\Gamma^2_{ij}&=&\frac{\delta^2\Gamma[\phi]}{ \delta\phi_i\delta\phi_j}|_{\phi=0}\,=\,-\,\frac{g^2}{ 4\pi^2} \int_{u,v}  \frac{(x-u)_i}{(x-u)^2}\frac{(y-v)_j}{(y-v)^2}\,
\langle\rho^a(u)\rho^b(v)\rangle_P\nonumber \\
\Gamma^3_{ijn}&=&\frac{\delta^3\Gamma[\phi]}{\delta\phi_i\delta\phi_j\delta\phi_n}|_{\phi=0}\,=\,-\,\frac{g^2}{8\pi^3} \int_{u,v,w} 
 \frac{(x-u)_i}{ (x-u)^2}\frac{(y-v)_j}{ (y-v)^2}\frac{(z-w)_n}{ (z-w)^2}\,
\langle\rho^a(u)\rho^b(v)\rho^c(w)\rangle_P\nonumber \\
\Gamma^4_{ijnm}&=&\,-\,\frac{g^2}{ 16\pi^4} \int_{u,v,w,w^\prime} 
 \frac{(x-u)_i}{(x-u)^2}\frac{(y-v)_j}{(y-v)^2}\frac{(z-w)_n}{(z^\prime-w)^2}\frac{(z^\prime-w^\prime)_m}{(z^\prime-w^\prime)^2}\,
\langle\rho^a(u)\rho^b(v)\rho^c(w)\rho^d(w^\prime)\rangle_P^{conn}
\end{eqnarray}
where the connected correlators are calculated as
\begin{eqnarray}
&&\langle\rho^a(u)\rho^b(v)\rangle_P\,=\,\int D[\rho]\,\rho^a(u)\rho^b(v)\,W_P[\rho]\nonumber \\
&&\langle\rho^a(u)\rho^b(v)\rho^c(w)\rangle_P\,=\,\int D[\rho]\,\rho^a(u)\rho^b(v)\rho^c(w)\,W_P[\rho]\nonumber \\
&&\langle\rho^a(u)\rho^b(v)\rho^c(w)\rho^d(w^\prime)\rangle_P^{conn}\,=\,\int D[\rho]\,\rho^a(u)\rho^b(v)\rho^c(w)\rho^d(w^\prime)\,W_P[\rho]|_{connected}
\end{eqnarray}
If the connected correlators of $\rho$ are small, these formulae can serve as a starting point for perturbative calculation of the entropy.

An alternative strategy could be to first integrate over $\phi$ in eq.(\ref{nonG}). This leads to the following expression:
\beq\label{rf}
tr[\hat \rho^N]= \int \prod_{\beta=1}^N D\rho_\beta \,\exp\left\{-\sum_{\beta=1}^N \left[ -\ln W_P[\rho_\beta]\,+\,\frac{1}{ 2\pi}
\int_{k,k^\prime} (b_{\beta+1}(k)-b_\beta(k))(b_{\beta+1}(k^\prime)-b_\beta(k^\prime)) \right]\right\}
\eeq
where $b_{N+1}=b_1$. This has the form of discrete lattice model with the same ``hopping term'' as in Section 2, but with a different on site potential $V[\rho]=-ln[W[\rho]]$.

To take the limit $N\rightarrow 1+\epsilon$ one would need to perform the integration over the replica fields analytically. If the projectile is dilute, the potential $V[\rho]$ is sharply peaked at small $\rho$ and the integral can be calculated in the ``tight binding approximation''. For the dense projectile on the other hand, the potential is small since large fluctuations of charge density are allowed. In this case one can expand around the free ``hopping term''. It would be interesting to study
the effects of non-Gaussianity on the entropy with the help of these approximations for realistic $W[\rho]$.

In this paper, in order to compute the entanglement entropy (and the entropy production), we have used the density matrix of soft gluons derived after tracing over valence gluons. As a further development of the above ideas, one can consider a further reduction of the density matrix. In particular, since
gluons with very small transverse momenta are unobservable, it could be interesting to trace over  gluons with momenta below some soft scale and consider the entropy of the remaining modes.  Another interesting venue would be to relate two-particle correlations, particularly long range rapidity correlations, to the concept of {\it mutual information}.  The latter is defined through  entanglement entropy of two different subsets of modes in the hadronic wave-function.
  

\section*{Acknowledgments}
We are thankful to Ramy Brustein for useful discussions. We also thank Dima Kharzeev for bringing reference \cite{momentum entropy} to our attention.
We thank the Physics Departments of the University of Connecticut and the Ben-Gurion University of the Negev for hospitality at times when this project was initiated and completed.
This research  was supported by the People Programme (Marie Curie Actions) of the European Union's Seventh Framework Programme FP7/2007-2013/ under REA  grant agreement \#318921; the DOE grant DE-FG02-13ER41989, the BSF grant \#2012124 and the  ISRAELI SCIENCE FOUNDATION grant \#87277111.

\end{document}